\documentclass[journal]{IEEEtran}

\usepackage[T1]{fontenc}
\usepackage[utf8]{inputenc}
\usepackage{lmodern}
\usepackage{microtype}

\usepackage{graphicx}
\usepackage[caption=false]{subfig}
\usepackage{enumerate}
\usepackage{physics}
\usepackage{amssymb}
\usepackage{hyperref}
\usepackage{interval}
\usepackage[capitalise]{cleveref}
\usepackage[group-separator={,}]{siunitx}
\usepackage[abbreviations,british]{foreign}
\usepackage{paralist}
\usepackage{pifont}
\usepackage{booktabs}

\intervalconfig{soft open fences}

\graphicspath{{figures/}}

\begin{document}

\title{Sifting Convolution on the Sphere}

\author{
	\IEEEauthorblockN{
		Patrick J. Roddy\IEEEauthorrefmark{1} and
		Jason D. McEwen\IEEEauthorrefmark{1}
	}

	\IEEEauthorblockA{
		\IEEEauthorrefmark{1}
		Mullard Space Science Laboratory (MSSL), University College London (UCL), Surrey RH5 6NT, UK 
		\newline
		\footnotesize \texttt{patrick.roddy.17@ucl.ac.uk}, \texttt{jason.mcewen@ucl.ac.uk}
	}
}

\maketitle

\begin{abstract}
	A novel spherical convolution is defined through the sifting property of the Dirac delta on the sphere.
	The so-called sifting convolution is defined by the inner product of one function with a translated version of another, but with the adoption of an alternative translation operator on the sphere.
	This translation operator follows by analogy with the Euclidean translation when viewed in harmonic space.
	The sifting convolution satisfies a variety of desirable properties that are lacking in alternate definitions, namely: it supports directional kernels; it has an output which remains on the sphere; and is efficient to compute.
	An illustration of the sifting convolution on a topographic map of the Earth demonstrates that it supports directional kernels to perform anisotropic filtering, while its output remains on the sphere.
\end{abstract}


\IEEEpeerreviewmaketitle{}

\section{Introduction}

\IEEEPARstart{M}{any} fields in science and engineering measure data on spherical manifolds, such as computer graphics~\cite{Ramamoorthi2004}, planetary science~\cite{Turcotte1981}, geophysics~\cite{Simons2006}, quantum chemistry~\cite{Choi1999}, cosmology~\cite{Bennett1996}, and computer vision~\cite{Cohen2018,Esteves2017,Cobb2020}.
Possible extensions to signal processing techniques developed in the Euclidean domain may be transferred to the spherical domain.
The convolution is an important signal processing technique between two signals defined on the \(2\)-sphere, which is central to filtering --- an integral part of spherical analyses.

Many definitions of spherical convolutions exist in the literature.
A spherical convolution operator would ideally exhibit a variety of desirable properties --- such a spherical convolution would accept directional inputs (\ie{} functions that are \emph{not} invariant under azimuthal rotation), whilst having the output remain on the sphere.
Moreover, the convolution would be efficient to compute.
Existing convolutions such as the \emph{isotropic convolution} (\eg{}~\cite{McEwen2007a,Wei2011,Kennedy2011}) and the \emph{left convolution}~\cite{Kennedy2011,Driscoll1994} restrict themselves to an axisymmetric kernel (\ie{} kernels that are invariant under azimuthal rotation).
The \emph{directional convolution} has an output which is not on the sphere (\eg{}~\cite{McEwen2007a,Wandelt2001}).
Lastly, the \emph{commutative anisotropic convolution}~\cite{Sadeghi2012,Khalid2012} and the \emph{directional convolution} are computationally demanding.
No existing spherical convolution satisfies all three desirable properties.

This letter presents an alternative spherical convolution, the \emph{sifting convolution}, defined through the sifting property of the Dirac delta --- in analogy to the Euclidean definition.
The convolution is anisotropic in nature and supports directional kernels.
The output remains on the sphere, even when both inputs are directional.
Moreover, the convolution is efficient to compute, and is commutative up to a complex conjugate.

The remainder of this letter is as follows.
\cref{sec:preliminaries} includes some mathematical preliminaries and reviews existing spherical convolutions in the literature.
\cref{sec:sifting_convolution} introduces the proposed sifting convolution.
\cref{sec:numerical_illustration} presents a demonstration of the convolution with a directional kernel.
Lastly, \cref{sec:conclusion} sets out some concluding remarks.

\section{Mathematical Background and Problem Formulation}\label{sec:preliminaries}

\subsection{Mathematical Preliminaries}

\subsubsection{Signals on the Sphere}

Consider a complex valued square-integrable function \(f(\omega)\) on the \(2\)-sphere \(\mathbb{S}^{2} = \{ \omega \in \mathbb{R}^{3} : \norm{\omega} = 1 \} \).
Here \(\omega=(\theta,\phi)\) parameterise a point on the unit sphere, where \(\theta \in \interval{0}{\pi}\) is the colatitude and \(\phi \in \interval[open right]{0}{2\pi}\) is the longitude.
The functions \(f(\omega)\) form the Hilbert Space \(L^{2}(\mathbb{S}^{2})\).
The complex inner product induces a norm \(\norm{f} = \sqrt{\braket{f}}\).
Signals on the sphere are functions with a finite induced norm.

\subsubsection{Spherical Harmonics}

The spherical harmonics are the complete orthonormal set of basis functions of the Hilbert space \(L^{2}(\mathbb{S}^{2})\).
By the completeness of spherical harmonics, any \(f \in L^{2}(\mathbb{S}^{2})\) may be decomposed as
\begin{equation}
	f(\omega) = \sum\limits_{\ell=0}^{\infty} \sum\limits_{m=-\ell}^{\ell} f_{\ell m} Y_{\ell m}(\omega),
\end{equation}
where \(f_{\ell m}\) are the spherical harmonic coefficients given by \(f_{\ell m} = \braket{f}{Y_{\ell m}}\).
The phase convention adopted here is \(Y_{\ell m}^{*} = {(-1)}^{m} Y_{\ell(-m)}\) such that \({f_{\ell m}^{*} = {(-1)}^{m} f_{\ell(-m)}}\) for a real field.
One often considers signals on the sphere with a bandlimit of \(L\), \ie{} signals such that \(f_{\ell m} = 0,\ \forall \ell \geq L\); and adopts the shorthand notation \(\sum_{\ell m} = \sum_{\ell=0}^{L-1} \sum_{m=-\ell}^{\ell}\).

\subsubsection{Dirac Delta}

The Dirac delta on the sphere satisfies the following normalisation and sifting properties, respectively: \(\int_{\mathbb{S}^{2}} \dd{\Omega(\omega)} \delta(\omega) = 1\) and \(\braket{\delta_{\omega'}}{f^{*}} = f(\omega')\), where \(\delta_{\omega'}(\omega)\) represents the Dirac delta rotated to some \(\omega'=(\theta',\phi')\).
The harmonic expansion of the Dirac delta is
\begin{equation}\label{eq:dirac_omega}
	\delta_{\omega'}(\omega) = \sum\limits_{\ell m} Y_{\ell m}^{*}(\omega') Y_{\ell m}(\omega),
\end{equation}
which follows trivially by the sifting property.

\subsubsection{Rotation of a Signal on the 2-Sphere}

The Euler angles may parameterise three-dimensional rotations with \(\rho = (\alpha,\beta,\gamma) \in \text{SO}(3)\), where \(\alpha \in \interval[open right]{0}{2\pi}\), \(\beta \in \interval{0}{\pi}\), and \(\gamma \in \interval[open right]{0}{2\pi}\).
The rotation operator \(\mathcal{R}_{\rho}\) consists of the sequence of rotations:
\begin{inparaenum}[(i)]
	\item \({\gamma}\) rotation about the \(z\)-axis;
	\item \({\beta}\) rotation about the \(y\)-axis; and
	\item \({\alpha}\) rotation about the \(z\)-axis.
\end{inparaenum}
The rotation of a function on the sphere is defined by \((\mathcal{R}_{\rho}f)(\omega) = f(\vb{R}_{\rho}^{-1} \omega)\), where \(\vb{R}_{\rho}\) is the three-dimensional rotation matrix corresponding to \(\mathcal{R}_{p}\).
The spherical harmonic coefficients of a rotated function read
\begin{equation}\label{eq:rotation_matrix}
	{(\mathcal{R}_{\rho} f)}_{\ell m} = \sum\limits_{m'=-\ell}^{\ell} D^{\ell}_{m'm}(\rho) f_{\ell m'},
\end{equation}
where \(D^{\ell}_{m'm}(\rho)\) are \emph{Wigner D matrices} which form the \(2\ell+1\)-dimensional representation of the rotation group for a given \({\ell}\).

\subsection{Spherical Convolutions}\label{sec:spherical_convolutions}

The conventional convolution between two functions on two-dimensional Euclidean space \(\mathbb{R}^{n}\) is
\begin{equation}
	(f \star g)(x) = \int\limits_{\mathbb{R}^{2}} \dd{y} f(x-y) g(y),
\end{equation}
where \(x,\ y \in \mathbb{R}^{n}\).
The convolution is commutative \({f \star g = g \star f}\).
A spherical counterpart of the convolution is required for functions defined on the sphere.
Alternative definitions of such a convolution exist in the literature but, while already useful, lack certain desirable properties.

The properties desired in the spherical extension of the convolution include:
\begin{inparaenum}[(i)]
	\item the support of directional kernels;
	\item an output which remains on the sphere; and
	\item efficient computation.
\end{inparaenum}
A convolution is considered computationally efficient here if its computational cost is no greater than the cost of fast spherical harmonic transforms, \ie{} \(\mathcal{O}(L^{3})\) (\eg{}~\cite{Driscoll1994,McEwen2011}).
Formulations of spherical convolutions exist in the literature but none satisfy all these properties.
A summary of existing spherical convolutions and their properties follows.

\subsubsection{Isotropic Convolution}

In real space the isotropic convolution (\eg{}~\cite{McEwen2007a,Wei2011,Kennedy2011}) is
\begin{equation}
	(f \odot g)(\omega) = \int\limits_{\mathbb{S}^{2}} \dd{\Omega(\omega')} f(\omega') {(\mathcal{R}_{\omega} g)}^{*} (\omega'),
\end{equation}
which in harmonic space becomes (\eg{}~\cite{McEwen2007a})
\begin{equation}\label{eq:isotropic_harmonic}
	{(f \odot g)}_{\ell m} = \sqrt{\frac{4\pi}{2\ell+1}} f_{\ell m} g_{\ell 0}^{*}.
\end{equation}
The isotropic convolution has the following properties:
\begin{inparaenum}[(i)]
	\item does not support directional kernels since \(g(\omega)\) must be axisymmetric;
	\item an output which remains on the sphere; and
	\item efficient computation since it is a product in harmonic space.
\end{inparaenum}

\subsubsection{Left Convolution}

The definition of the left convolution~\cite{Kennedy2011,Driscoll1994} in real space is
\begin{equation}
	(f \circleddash g)(\omega) = \int\limits_{\text{SO}(3)} \dd{\rho(\rho)} f(\rho\eta) g(\rho^{-1}\omega),
\end{equation}
where \({\eta}\) is the north pole, and \(\dd{\rho(\rho)}=\sin{\beta} \dd{\alpha} \dd{\beta} \dd{\gamma}\) is the usual invariant measure on \(\text{SO}(3)\).
The harmonic representation of this convolution is
\begin{equation}
	{(f \circleddash g)}_{\ell m} = 2\pi \sqrt{\frac{4\pi}{2\ell+1}} f_{\ell m} g_{\ell0},
\end{equation}
As the harmonic representations suggest, the isotropic and left convolutions are closely related, as elaborated in~\cite{Kennedy2011}.
Hence, the properties are similar.
The left convolution has the following properties:
\begin{inparaenum}[(i)]
	\item does not support directional kernels since \(g(\omega)\) must be axisymmetric;
	\item an output which remains on the sphere; and
	\item efficient computation since it is a product in harmonic space.
\end{inparaenum}

\subsubsection{Directional Convolution}

Rotations on the sphere are the spherical counterpart of translations in the Euclidean domain in real space.
Hence, the standard directional convolution is
\begin{equation}\label{eq:directional_convolution}
	(f \circledast g)(\rho) = \int\limits_{\mathbb{S}^{2}} \dd{\Omega(\omega)} f(\omega) {(\mathcal{R}_{\rho} g)}^{*} (\omega).
\end{equation}
Upon expanding in harmonic space, this becomes (\eg{}~\cite{McEwen2007a,Wandelt2001})
\begin{equation}
	(f \circledast g) (\rho) = \sum\limits_{\ell m} \sum\limits_{m'=-\ell}^{\ell} f_{\ell m} {\left( D^{\ell}_{m'm}(\rho) g_{\ell m'} \right)}^{*},
\end{equation}
and hence, the output is on \(\text{SO}(3)\).
Fast algorithms exist~\cite{McEwen2007a,Wandelt2001,Wiaux2007,McEwen2013c} but the convolution remains less efficient than a spherical harmonic transform.
The directional convolution has the following properties:
\begin{inparaenum}[(i)]
	\item does support directional kernels;
	\item an output which does not remain on the sphere due to the 3D rotation of the kernel; and
	\item expensive computation.
\end{inparaenum}

\subsubsection{Commutative Anisotropic Convolution}

The definition of the commutative anisotropic convolution~\cite{Sadeghi2012,Khalid2012} is
\begin{equation}
	(f \oplus g)(\omega) = \int\limits_{\mathbb{S}^{2}} \dd{\Omega(\omega')} {(\mathcal{R}_{(\phi,\theta,\pi-\phi)} f)}(\omega') g(\omega'),
\end{equation}
which on expansion reads
\begin{equation}
	(f \oplus g)(\omega) = \sum\limits_{\ell m} \sum\limits_{m'=-\ell}^{\ell} D^{\ell}_{m'm}(\phi,\theta,\pi-\phi) f_{\ell m'} g_{\ell m}^{*}.
\end{equation}
The limitation here is that one must specify the initial rotation as \({\gamma=\pi-\alpha}\) in order for the convolution to be commutative.
The complexity of the convolution is \(\mathcal{O}(L^{3}\log{L})\), and hence, it is less efficient than a spherical harmonic transform.
The convolution has the following properties:
\begin{inparaenum}[(i)]
	\item it supports directional kernels;
	\item an output which remains on the sphere; and
	\item expensive computation.
\end{inparaenum}

\subsection{Problem Formulation}

\cref{tab:properties} presents a summary of the spherical convolutions discussed and their properties.
No existing definition of a spherical convolution has all the desired properties discussed in \cref{sec:spherical_convolutions}.
In this work, the sifting convolution, which satisfies all desirable properties, is presented.

\begin{table}
	\centering
	\caption{
		Properties of spherical convolutions.
	}\label{tab:properties}
	\begin{tabular}{@{}rccc@{}}
		\toprule
		                        & Anisotropic & \(\mathbb{S}^{2}\) Output & Efficient \\
		\midrule
		Isotropic               & \ding{55}   & \ding{51}                 & \ding{51} \\
		Left                    & \ding{55}   & \ding{51}                 & \ding{51} \\
		Directional             & \ding{51}   & \ding{55}                 & \ding{55} \\
		Commutative Anisotropic & \ding{51}   & \ding{51}                 & \ding{55} \\
		Sifting (this work)     & \ding{51}   & \ding{51}                 & \ding{51} \\
		\bottomrule
	\end{tabular}
\end{table}

\section{Sifting Convolution}\label{sec:sifting_convolution}

This work defines the sifting convolution which permits directional kernels, whose output remains on the sphere and is efficient to compute.
Moreover, it is commutative up to a complex conjugate.
The sifting convolution is constructed using a novel translation operator defined on the sphere.

\subsection{Translation Operator}\label{sec:translation_operator}

In real space, the rotation operator on the sphere is the usual analogue of the translation operator in the Euclidean setting.
One may define an alternative operator, \(\mathcal{T}_{\omega}\), which follows as the analogue of the Euclidean setting but in harmonic space.
This translation is in contrast to the standard rotation as it considers two angles rather than three and thereby its output remains on the sphere.
In practice the translation operator is defined as a product of basis functions.
In the Euclidean setting, \eg{} \(\mathbb{R}\), the complex exponentials \(\phi_{u}(x) = \exp{iux}\), with \(x,\ u \in \mathbb{R}\) form the standard orthonormal basis.
A shift of coordinates defines the translation of the basis functions: \(\phi_{u}(x + x') = \phi_{u}(x') \phi_{u}(x)\), with \(x' \in \mathbb{R}\) and where the final equality follows by the standard rule for exponents.
The definition of the translation of the spherical harmonics on the sphere follows by analogy with the representation as a product of basis functions:
\begin{equation}
	(\mathcal{T}_{\omega'} Y_{\ell m})(\omega) \equiv Y_{\ell m}(\omega') Y_{\ell m}(\omega),
\end{equation}
where \(\omega'=(\theta',\phi')\).

This leads to a natural harmonic expression for the translation of a general arbitrary function \(f \in L^{2}(\mathbb{S}^{2})\)
\begin{equation}\label{eq:translation_real}
	(\mathcal{T}_{\omega'} f)(\omega) = \sum\limits_{\ell m} f_{\ell m} Y_{\ell m}(\omega') Y_{\ell m}(\omega),
\end{equation}
implying
\begin{equation}\label{eq:translation_harmonic}
	{(\mathcal{T}_{\omega'} f)}_{\ell m} = f_{\ell m} Y_{\ell m}(\omega').
\end{equation}
This translation operator is considered further in \cref{sec:translation_interpretation} to build greater intuition.

\subsection{Convolution Operator}

With a translation operator to hand, one may define the sifting convolution on the sphere of \(f,\ g \in L^{2}(\mathbb{S}^{2})\) in the usual manner by the inner product
\begin{equation}\label{eq:convolution_real}
	(f \circledcirc g)(\omega) \equiv \braket{\mathcal{T}_{\omega}f}{g},
\end{equation}
noting the use of the alternative translation operator defined in \cref{sec:translation_operator}.

In harmonic space this simplifies to the product
\begin{equation}\label{eq:convolution_harmonic}
	{(f \circledcirc g)}_{\ell m} = f_{\ell m} g_{\ell m}^{*},
\end{equation}
as
\begin{align}
	(f & \circledcirc g)(\omega) = \braket{\mathcal{T}_{\omega}f}{g} \nonumber{}                                                                                                                                  \\
	   & = \int\limits_{\mathbb{S}^{2}} \dd{\Omega(\omega')} (\mathcal{T}_{\omega}f)(\omega') g^{*}(\omega') \nonumber{}                                                                                          \\
	   & = \int\limits_{\mathbb{S}^{2}} \dd{\Omega(\omega')} \sum\limits_{\ell m} f_{\ell m} Y_{\ell m}(\omega') Y_{\ell m}(\omega) \sum\limits_{\ell' m'} g_{\ell' m'}^{*} Y_{\ell' m'}^{*}(\omega') \nonumber{} \\
	   & = \sum\limits_{\ell m} f_{\ell m} g_{\ell m}^{*} Y_{\ell m}(\omega).
\end{align}
Since the harmonic representation of the convolution is simply a product (again by analogy with the harmonic representation of the Euclidean convolution), it is efficient to compute.
Note that harmonic multiplication has been considered before~\cite{Kennedy2011}; although it has been used here to define a new anisotropic convolution operator, introducing a conjugation and elaborating a real space interpretation.

\subsection{Translation Interpretation}\label{sec:translation_interpretation}

One may show that the translation operator is simply a (sifting) convolution of a function with the shifted Dirac delta function:
\begin{align}
	(f \circledcirc \delta_{\omega'})(\omega)
	 & = \sum\limits_{\ell m} f_{\ell m} Y_{\ell m}(\omega') Y_{\ell m}(\omega) \nonumber{} \\
	 & = (\mathcal{T}_{\omega'}f)(\omega),
\end{align}
by noting \cref{eq:convolution_harmonic} and where the final equality follows by \cref{eq:translation_real}.
The sifting convolution and translation are thus natural analogues of the respective operators defined in Euclidean space.

\subsection{Properties}

The sifting convolution has all the desired properties discussed in \cref{sec:spherical_convolutions}, namely, the convolution accepts directional inputs, has an output which remains on the sphere, and is efficient to compute.
\cref{tab:properties} summarises the properties of the sifting convolution and compares them to the properties of alternative spherical convolutions.

The translation preserves symmetries, which means that any symmetry that exists in the initial kernel definition will be present after the translation.
Thus, one must be careful when choosing a kernel for a convolution to ensure it has the desired properties when translated, \eg{} spatial localisation.
To perform anisotropic smoothing that is localised (the usual interpretation), the translated kernel also needs to be localised.
If the kernel has, say, even azimuthal symmetry, when it is translated to \(\omega'=(\theta', \phi')\) it will have a localised component both at \(\phi'\) and \(-\phi'\).
While this is not a problem \emph{per se}, for the usual interpretation of smoothing one would desire a localised component at \(\phi'\) only.
This can be achieved by ensuring the original kernel does not exhibit a symmetry that would lead to multiple localised components once translated.
The \emph{harmonic Gaussian} introduced in \cref{sec:numerical_illustration} satisfies the desired property.

\section{Numerical Illustration}\label{sec:numerical_illustration}

This section demonstrates the effect of the sifting convolution through the application of a directional kernel to an example signal on the sphere.

Define the \emph{harmonic Gaussian} as a two-dimensional Gaussian in harmonic space by
\begin{equation}
	f_{\ell m} = \exp(-\left(\frac{{\ell}^{2}}{2\sigma_{\ell}^{2}} + \frac{{m}^{2}}{2\sigma_{m}^{2}}\right)).
\end{equation}
In effect, this function is the standard axisymmetric Gaussian in \(\ell{}\) modulated by a Gaussian in \(m\).
Note this function is not real --- if required one can define only the positive \(m\) components and impose reality by the conjugate symmetry relationship in harmonic space.
The function is directional, and hence, is useful in illustrating the effect of the sifting convolution on the sphere.
All later computations use the \texttt{SSHT}\footnote{\url{http://astro-informatics.github.io/ssht/}} code~\cite{McEwen2011}.

\begin{figure}
	\centering
	\subfloat[\(\Re{(\mathcal{T}_{\omega'}f_{A})(\omega)}\)]{\includegraphics[trim={22 6 7 1},clip,width=.5\columnwidth]{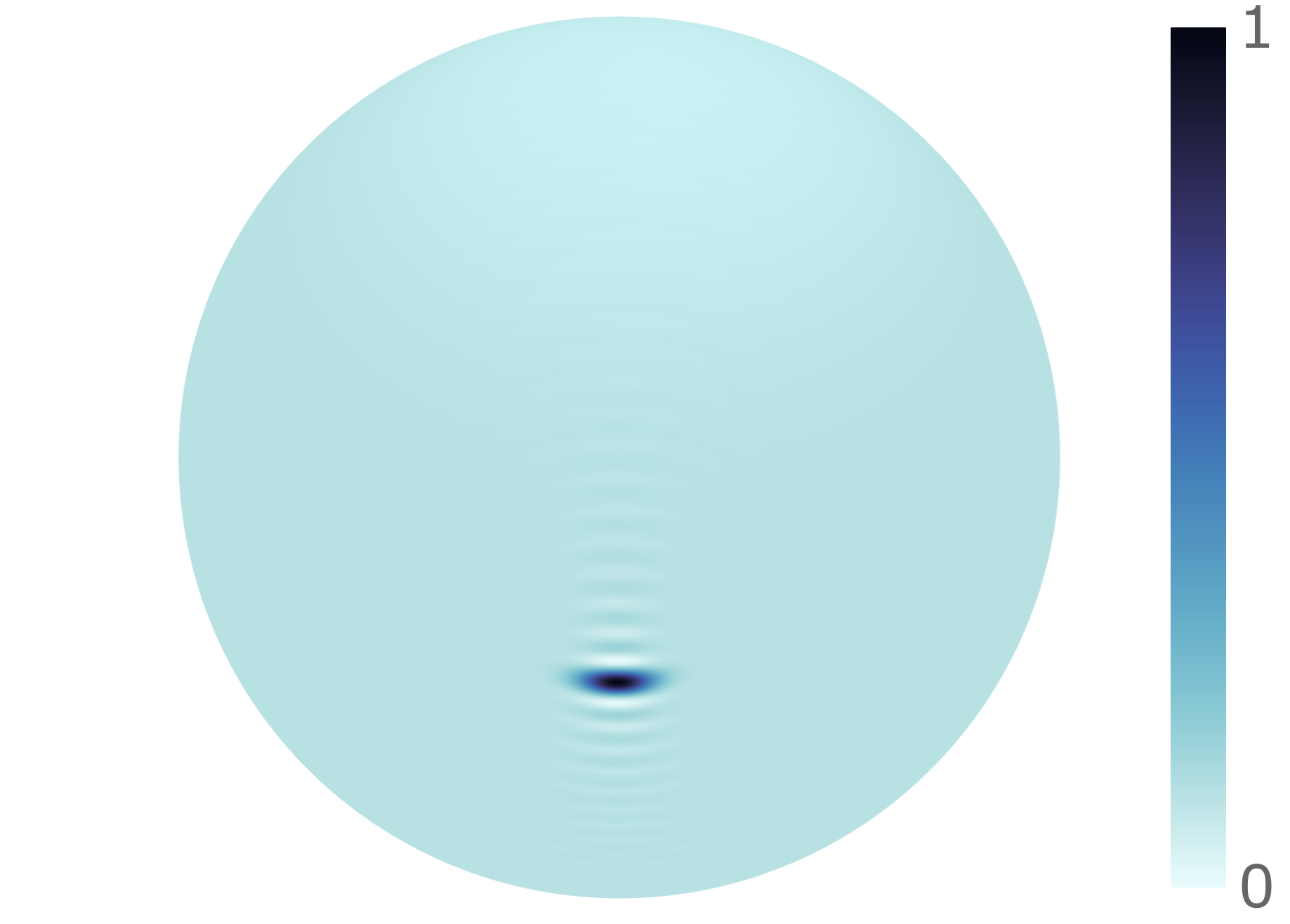}}
	\hfill
	\subfloat[\(\Re{(\mathcal{T}_{\omega'}f_{B})(\omega)}\)]{\includegraphics[trim={22 6 7 1},clip,width=.5\columnwidth]{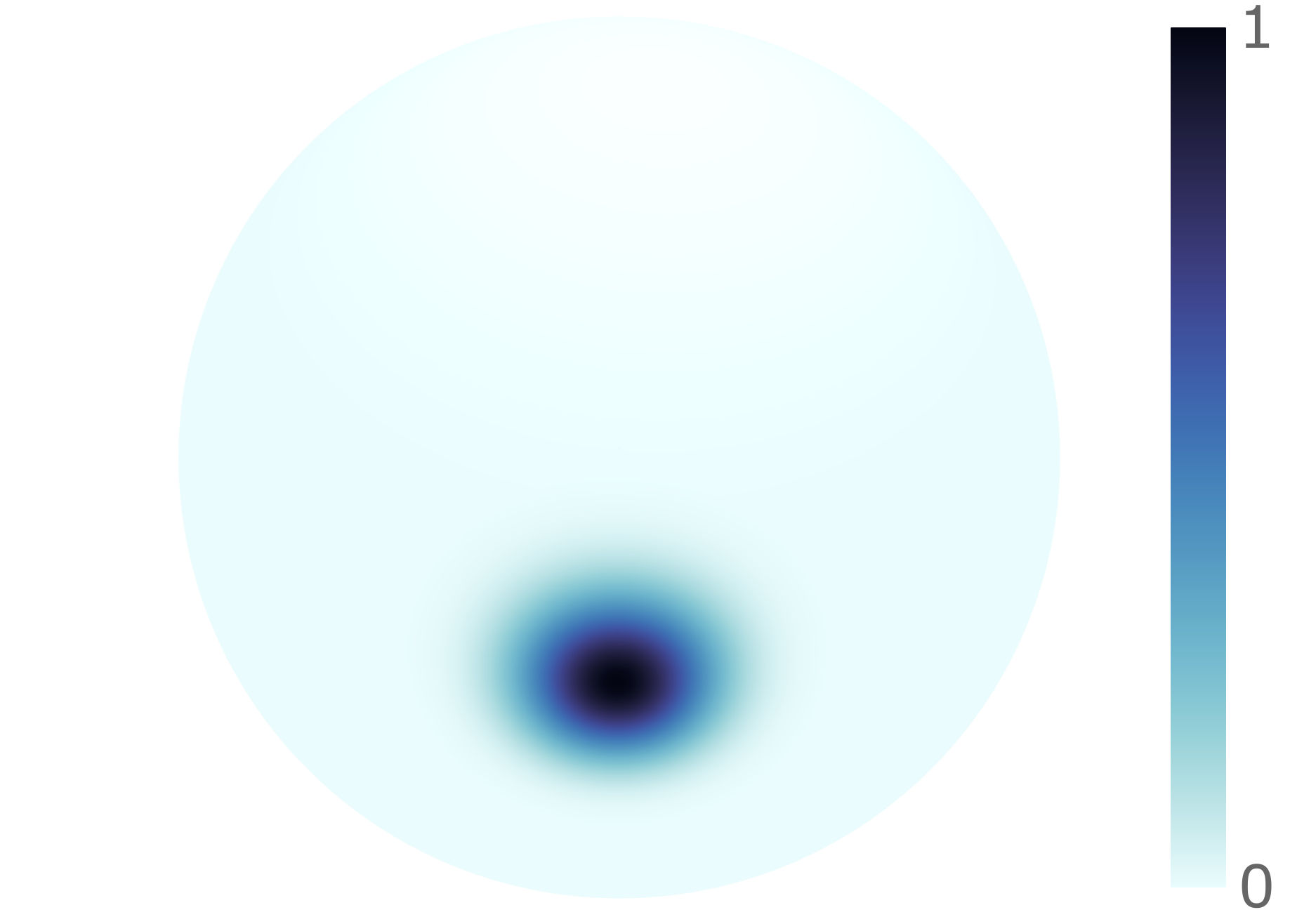}}
	\caption{
		A harmonic Gaussian translated to some \(\omega'=(\theta',\phi')\) (bandlimited at \(L=128\)).
		Panel (a) corresponds to a more elongated kernel \(f_{A}\), where \((\sigma_{\ell},\sigma_{m}) = (10^{2}, 10^{1})\); whereas panel (b) corresponds to a more symmetric kernel \(f_{B}\), where \((\sigma_{\ell},\sigma_{m}) = (10^{1}, 10^{1})\).
		The colour is between zero and one, reflecting the scaled intensity of the field.
	}\label{fig:translated}
\end{figure}

Consider two differently sized harmonic Gaussians on the sphere to see the effect on the sifting convolution.
\cref{fig:translated} shows both an elongated (left panel) and symmetric (right panel) translated harmonic Gaussian.

\begin{figure}
	\centering
	\includegraphics[trim={22 6 7 1},clip,width=.5\columnwidth]{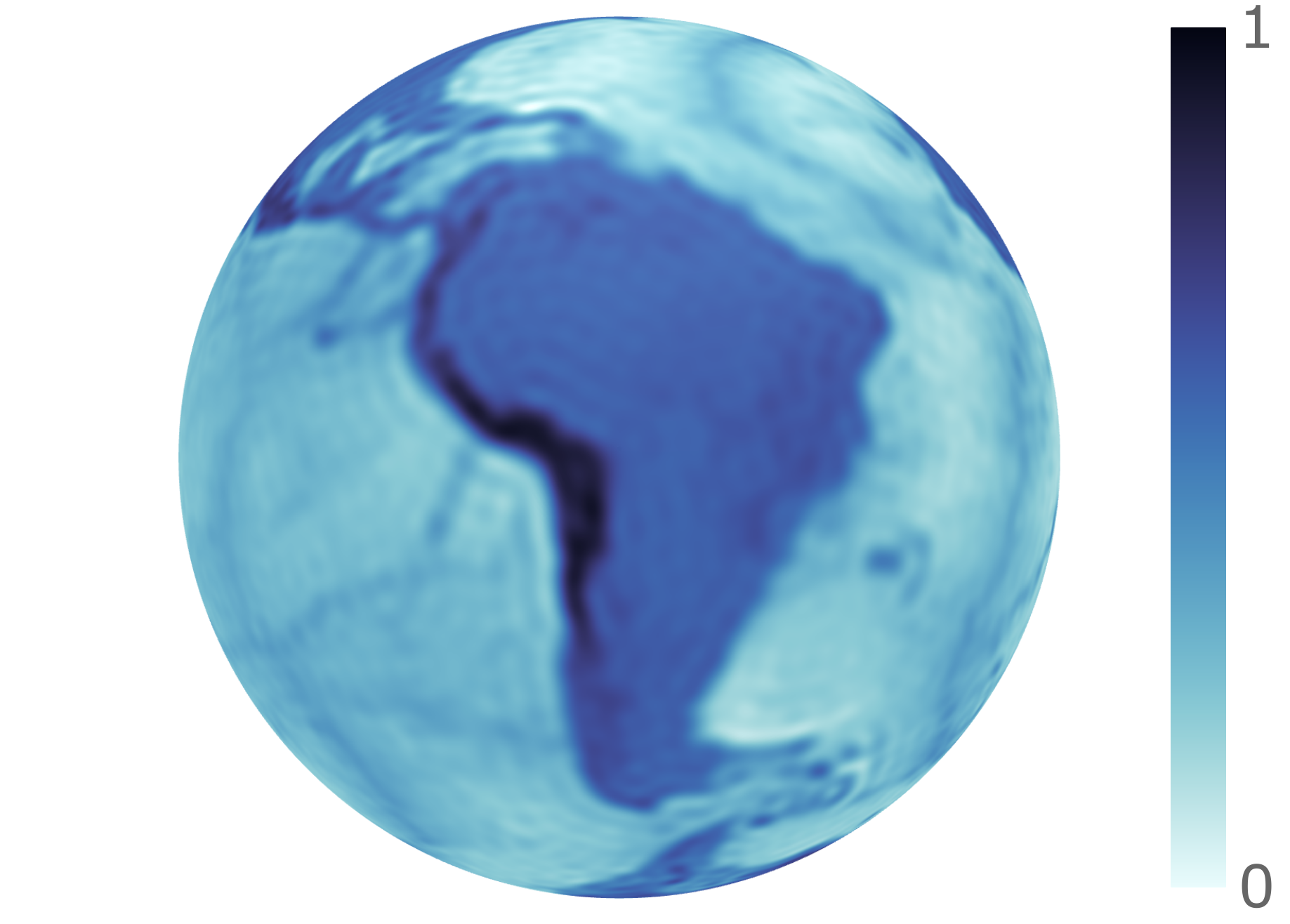}
	\caption{
		EGM2008 dataset centred on a view of South America (bandlimited at \(L=128\)).
		The colour is between zero and one, reflecting the scaled intensity of the field.
	}\label{fig:earth}
\end{figure}

\begin{figure}
	\centering
	\subfloat[\(\Re{(f_{A} \circledcirc g)(\omega)}\)]{\includegraphics[trim={22 6 7 1},clip,width=.5\columnwidth]{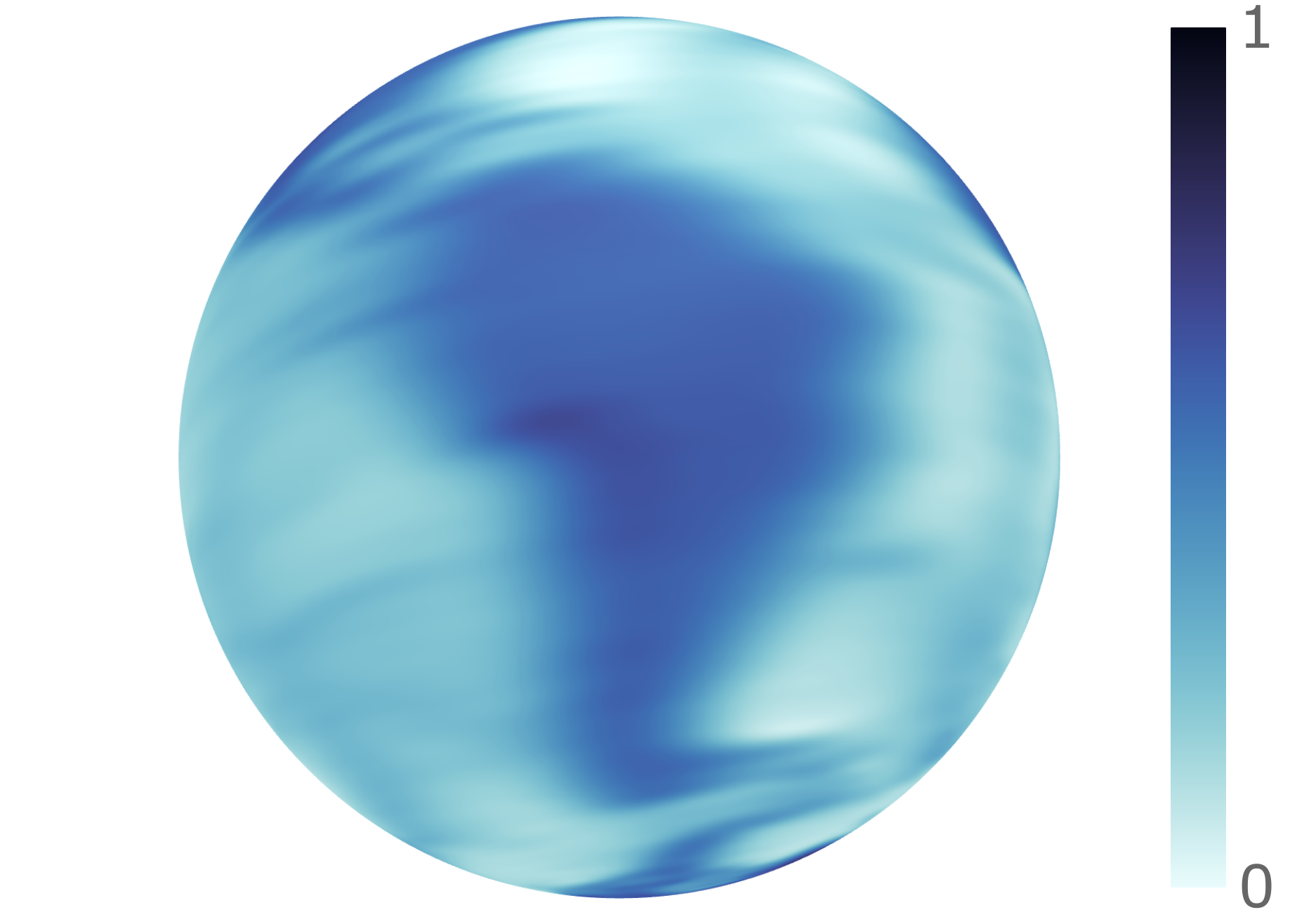}}
	\hfill
	\subfloat[\(\Re{(f_{B} \circledcirc g)(\omega)}\)]{\includegraphics[trim={22 6 7 1},clip,width=.5\columnwidth]{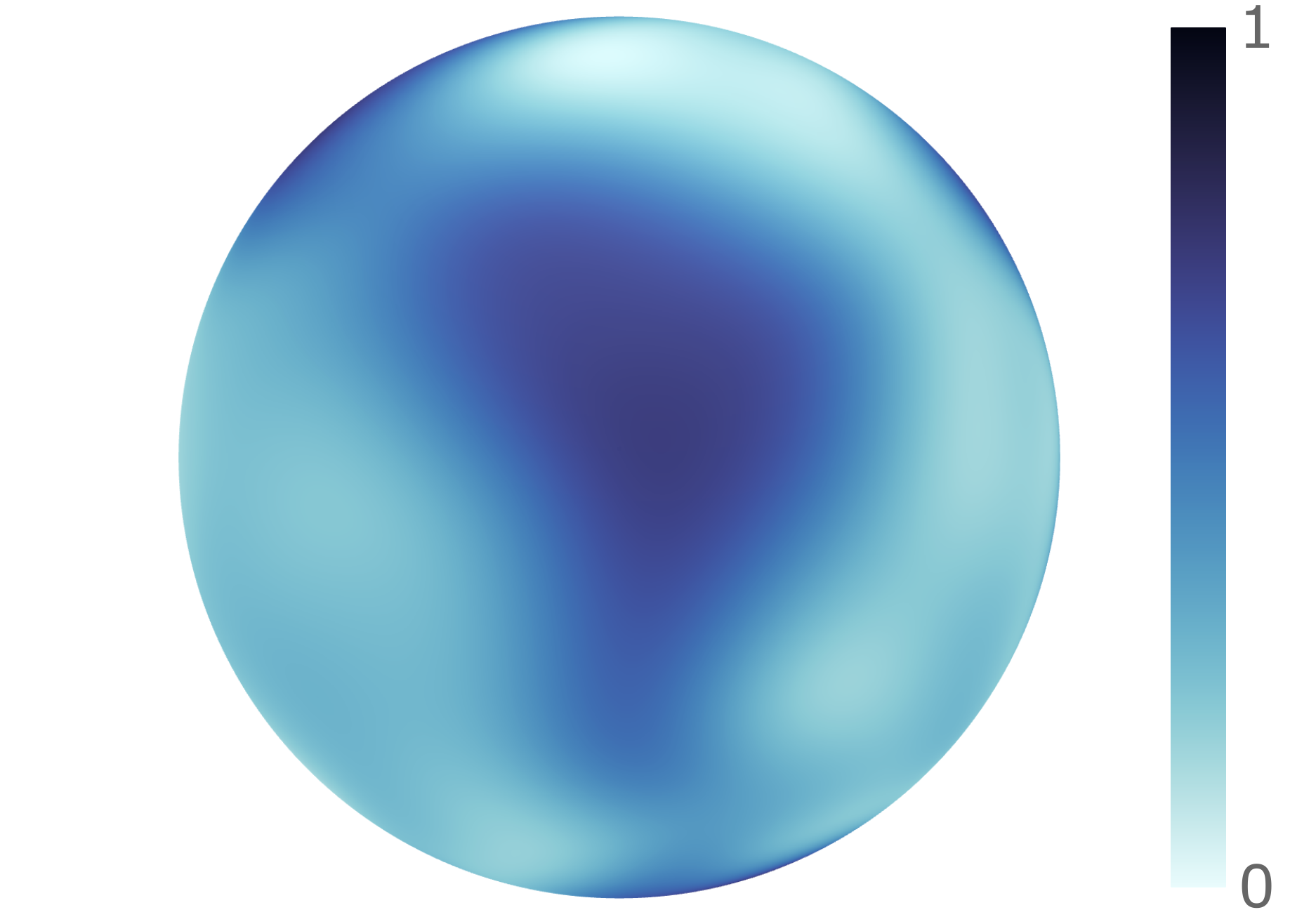}}
	\caption{
		The real part of the sifting convolution between the EGM2008 dataset and the harmonic Gaussian, rotated to view of South America (bandlimited at \(L=128\)).
		Panel (a) corresponds to a more elongated kernel \(f_{A}\), where \((\sigma_{\ell},\sigma_{m}) = (10^{2}, 10^{1})\); whereas panel (b) corresponds to a more symmetric kernel \(f_{B}\), where \((\sigma_{\ell},\sigma_{m}) = (10^{1}, 10^{1})\).
		As expected, the resultant sifting convolved Earth map exhibits greater anisotropic smoothing in panel (a) than in panel (b).
		It is clear that the sifting convolution supports directional kernels to perform anisotropic filtering (smoothing), while the output remains on the sphere.
		The colour is between zero and one, reflecting the scaled intensity of the field.
	}\label{fig:convolved}
\end{figure}

To study the effect of the sifting convolution, consider the \emph{Earth Gravitational Model EGM2008} dataset~\cite{Pavlis2013}.
This dataset is the topographic map of the Earth.
\cref{fig:earth} presents the dataset up to an order of \(L=128\).
The sifting convolution is then performed between the Earth representation and the harmonic Gaussian with the resultant plot given in \cref{fig:convolved}.
As expected, when the elongated kernel is considered, as shown in the left panel, the result exhibits greater anisotropic smoothing than when considering the symmetric kernel, as shown in the right panel.
It is clear that the sifting convolution supports directional kernels to perform anisotropic filtering (smoothing), while the output remains on the sphere.

\section{Conclusion}\label{sec:conclusion}

This work presents the \emph{sifting convolution} on the sphere and demonstrates its application.
The convolution accepts directional functions as inputs, has an output which remains on the sphere, and is efficient to compute.
The sifting convolution is defined in the usual manner through the inner product but with an alternative translation operator on the sphere.
This follows by analogy with the Euclidean translation when viewed as a convolution with a shifted Dirac delta function.
An illustration of the sifting convolution on the topographic map of the Earth demonstrates that it supports directional kernels to perform anisotropic filtering, while its output remains on the sphere.
Convolutions are an important part of signal processing techniques, hence, the sifting convolution can play an integral role in constructions of alternate spherical analysis techniques, which is the focus of future work.

\pagebreak
\bibliographystyle{IEEEtran}
\bibliography{library}

\end{document}